\documentclass[%
 reprint,
 nofootinbib,
 amsmath,amssymb,
pra,
]{revtex4-2}

\usepackage{physics}
\usepackage{graphicx}
\usepackage{dcolumn}
\usepackage{bm}
\usepackage{hyperref}

\begin{document}

\preprint{APS/123-QED}

\title{An optomechanical discrete variable quantum teleportation scheme}

\author{Samuel Pautrel}
\author{Zakari Denis}%
\author{J\'er\'emy Bon}%
\author{Adrien Borne}%
\author{Ivan Favero}%
 \email{ivan.favero@u-paris.fr}
\affiliation{%
 Mat\'eriaux et Ph\'enom\`enes Quantiques, Universit\'e de Paris, CNRS UMR 7162, 10 rue Alice Domon et L\'eonie Duquet, 75013 Paris, France
}%

\date{\today}

\begin{abstract}
We propose an experimental protocol to realize discrete variable quantum teleportation using optomechanical devices. The photonic polarization superposition state of a single photon is teleported to a phononic superposition of two micromechanical oscillators by means of photon/phonon entanglement generation and optical Bell state measurement using two-photon interference. Verification of the protocol is performed by coherent state transfer between the mechanical devices and light. Simulations show the feasibility of the proposed scheme at millikelvin temperatures using state-of-the-art gigahertz optomechanical devices. 
\end{abstract}

\maketitle


\section{Introduction}

Quantum teleportation \cite{Bennett93} consists in transferring ("teleporting") an arbitrary quantum state between two objects that are possibly distinct in nature, and possibly distant from one another. Quantum teleportation protocols always rely on entanglement, a ressource at the heart of several quantum technologies, and they have been implemented in a variety of systems: first with optical photons \cite{Zeil97,Furusawa98}, then amongst distinct atoms of a molecule \cite{Niel98} or of a linear trap \cite{Blatt04, Wineland04,Olmschenk09}, and more recently from a photon to a solid-state spin \cite{Gao13} and between two solid-state qubits \cite{Hanson14,Chou18}. In an effort to extend these teleportation principles to a macroscopic scale, protocols were theoretically proposed to transfer quantum states between light and vibrating mechanical systems, in the case of continuous variables \cite{Manc03, Pirandola03, Hofer11, Feli17}. However to date there is no report of optomechanical quantum teleportation following these propositions, and an experimental gap remains. The emergence of nanoscale gigahertz optomechanical resonators \cite{Painter04, Favero10,Favero11} may allow to fill this gap, as suggested by recent experiments in the quantum realm such as the optical entanglement of mechanical systems \cite{Ried18} and the test of related Bell inequalities \cite{Marinkovic18}.

In this article, we introduce a discrete-variable optomechanical quantum teleportation scheme compatible with current technology. In the present scheme, the state of a single photon polarization-encoded qubit is teleported onto a dual-rail phononic qubit encoded in two mechanical resonators. Our approach relies on measurement induced entanglement in the experimentally relevant weak-coupling and good-cavity (resolved-sideband) regime, with mechanical systems originally cooled close to their ground state. We analytically and numerically model the full protocol, and quantify its resilience against various sources of noise and disorder. The calculated efficiencies and fidelities, which can exceed 0.9 with realistic parameters, indicate that storing the quantum information conveyed by a single photon within a long-lasting mechanical excitation is now an objective at reach.



\section{Principle}
\begin{figure}[ht!]
    \centering
    \includegraphics[width=0.5\textwidth]{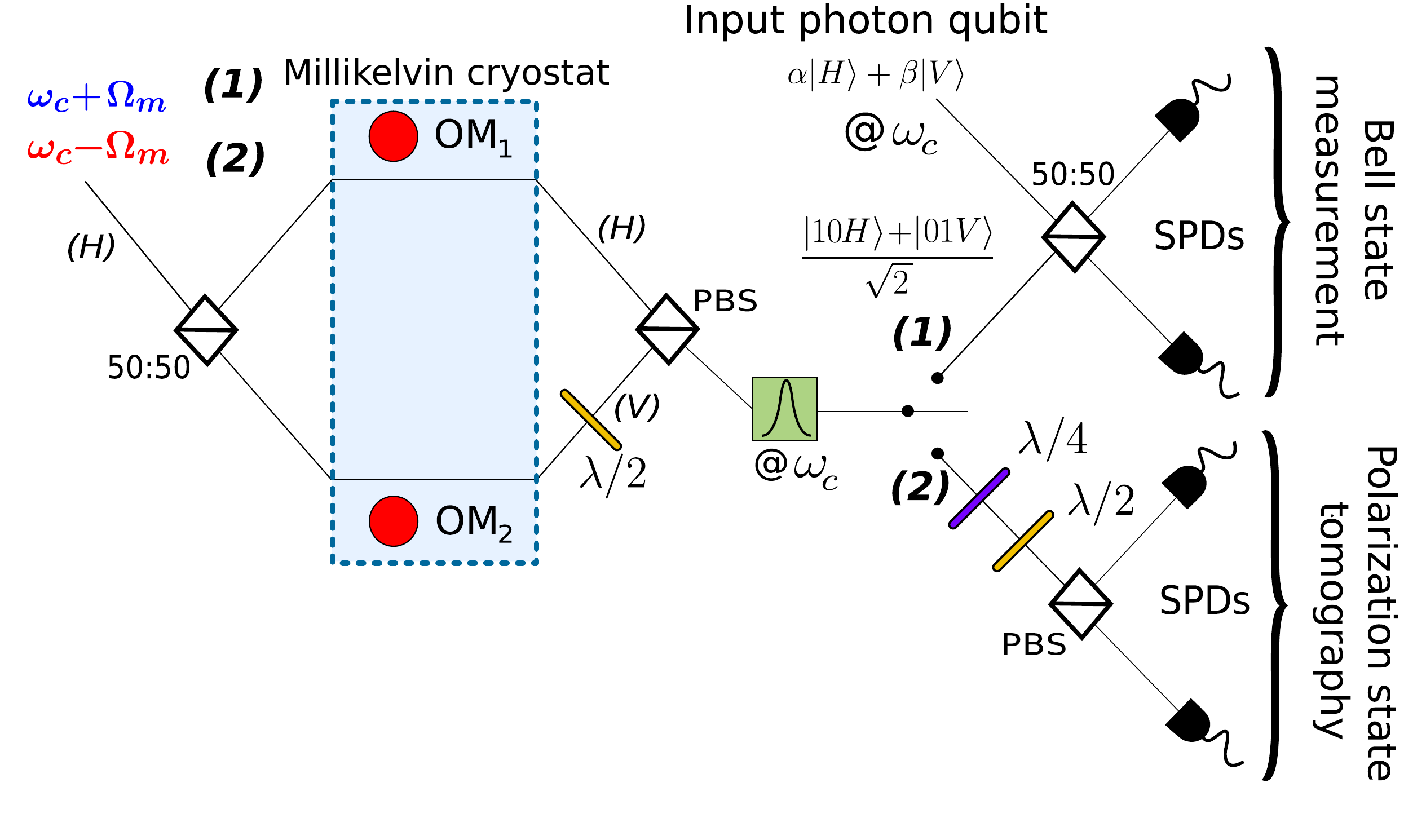}
    \caption{Schematics of the protocol setup with its two steps. A blue-detuned pulse \textbf{(1)} probabilistically generates a phonon in one of the two resonators, entangling them with the output optical mode. After filtering out the pump tone, we interfere the output photon with an arbitrarily polarized, but otherwise indistinguishable, photon. A coincidence event on the two SPDs heralds the teleportation (see text for details). A red-detuned pulse \textbf{(2)} coherently maps the mechanical state onto a photon whose polarization state is subsequently analyzed by means of quantum optical tomography.}
    \label{fig:scheme}
\end{figure}

The scheme we propose uses two identical optomechanical resonators with an optical resonant frequency $\omega_c$ coupled to a mechanical mode of angular frequency $\Omega_m$ with a single-photon coupling rate $g_0$, placed at millikelvin temperatures in order to initially reside in their mechanical ground state. The protocol setup is depicted in Fig.~\ref{fig:scheme}. Each resonator is placed in one arm of a balanced Mach-Zehnder interferometer. The devices are driven with a linearly polarized (say, horizontally polarized $H$) laser and the light at the output of the two arms is recombined on a polarizing beamsplitter (PBS) after rotation of one arm's polarization with a half-wave plate ($\lambda/2$). The resulting light is bandpass filtered with a high-finesse Fabry-Perot cavity in order to select the photons scattered at the resonant frequency $\omega_c$ and then directed to one of the two measurement units involving single-photon detectors (SPD), which respectively implement Bell state measurement and quantum state tomography. The protocol we present is reminiscent of a former protocol involving ions \cite{Olmschenk09}, and starts with both mechanical resonators in their ground state ($n_1=n_2=0$, where $n$ is the phonon number). It consists in two steps: \textbf{(1)} teleportation of the quantum information from a photonic qubit to a phononic qubit, and \textbf{(2)} optical readout of the teleported state. 

In \textbf{(1)} a blue-detuned H-polarized pulse (frequency $\omega_+ = \omega_c + \Omega_m$) generates a phonon in one of the two optomechanical devices $\text{OM}_1$ or $\text{OM}_2$ with low probability \textit{via} a cavity-enhanced optomechanical (Raman) interaction (Stokes process). The creation of such a delocalized mechanical excitation is concomitant with the presence of one photon at the cavity frequency $\omega_c$ at the output of the interferometer, after filtering the much stronger blue-detuned pump tone \cite{Ried18}. In order to erase the which-path information, the optical cavity frequencies must be as close as possible to maximize the indistinguishability of their output photons. At this point and in the ideal case, the two mechanical resonators are entangled with the optical output of the Mach-Zehnder in a state proportional to $\ket{01V}+\ket{10H}$, where $\ket{n_1n_2P}$ denotes the product state with $n_1$ phonons in $\text{OM}_1$, $n_2$ phonons in $\text{OM}_2$ and a photon with polarization $P=H$ or $V$.

We then perform a two-photon interference experiment on a $50:50$ beam-splitter between the previous scattered photon and an arbitrarily polarized single-photon whose state $\ket{\Phi}= \alpha\ket{H}+\beta\ket{V}$ is to be teleported. Assuming these two photons are spectrally indistinguishable, a coincidence on the two SPDs at the output ports of the beamsplitter indicates the projection of the bipartite photonic state in the antisymmetric Bell state $\ket{\Psi^-}= \frac{1}{\sqrt{2}}(\ket{HV}-\ket{VH})$ before the beam-splitter. This is a partial canonical Bell state measurement (BSM). Thus a coincidence event following this blue-detuned pulse heralds the teleportation of arbitrary amplitudes $\alpha$ and $\beta$ into a bipartite dual-rail phononic state encoded on the two mechanical resonators $1$ and $2$, $\ket{\Psi^{rot}_{mech}} = \beta\ket{10}-\alpha\ket{01}$ \textbf{(1)}. 

To retrieve the stored information, we apply \textbf{(2)} a red-detuned pulse (frequency $\omega_-=\omega_c-\Omega_m$) to implement a beamsplitter interaction through cavity-enhanced optomechanical (Raman) scattering (anti-Stokes) and coherently map the mechanical state onto a single photon at $\omega_c$, whose polarization state is then analyzed by conventional optical quantum state tomography. If not analyzed, the obtained photon could be guided to a remote similar experiment in order to perform teleportation again, propagating the quantum information. The mechanical resonators would then act as quantum memories \cite{Wall19}, an approach that appears relevant now that nanomechanical dissipation mechanisms are well understood \cite{Hamoumi18} and promise $Q \times f$ factors (mechanical quality factor times frequency) approaching $10^{20}$ \cite{Hamoumi18,MacCabe19}.

\section{Model}

\subsubsection{General model and evolution operators}

We start from the Hamiltonian for a driven-dissipative optomechanical system with initial mechanical occupancy $n_0 \ll 1$, resonant cavity frequency $\omega_c$, optical linewidth $\kappa$ dominated by the extrinsic decay rate $\kappa_e$, mechanical frequency $\Omega_m$ and mechanical damping rate $\gamma_m$. The optical cavity is driven with a laser at frequency $\omega_\pm = \omega_c \pm \Omega_m$. The total Hamiltonian reads:
\begin{subequations}
 \label{equ:Hamiltonian}
 \begin{align}
    \hat{H} &= \hat{H}_0+\hat{H}_{int}+\hat{H}_{env}+\hat{H}_{dr\pm} \\
    \hat{H}_0 &= \hbar\omega_c\hat{a}^\dag\hat{a} + \hbar \Omega_m \hat{b}^\dag\hat{b}\\
    \hat{H}_{int} &= -\hbar g_0 \hat{a}^\dag\hat{a} (\hat{b}^\dag+\hat{b})\\
    \hat{H}_{dr\pm} &= \hbar(s_\pm^*e^{i\omega_\pm t}\hat{a}+s_\pm e^{-i\omega_\pm t}\hat{a}^\dag)
\end{align}
\end{subequations}
where $\hat{H}_{int}$ and $\hat{H}_{env}$ account for the optomechanical interaction and for the environment respectively, and $\hat{H}_{dr\pm}$ is the driving term with $\abs{s_\pm}=\sqrt{\kappa P_\pm / \hbar \omega_\pm}$ the incoming photon flux for a laser power $P_\pm$. $\hat{H}_{0}$ is the bare Hamiltonian for the two harmonic oscillators where $\hat{a}$ ($\hat{b}$) is the annihilation operator for the optical (mechanical) mode under consideration.
Starting from the Heisenberg picture, we rotate our operators with respect to $\hat{H}_0$, following \cite{Woolley13}. We then assume to be in the good-cavity (resolved-sideband) regime, and consider interaction times $\tau$ shorter than $2\pi(n_0\gamma_m)^{-1}$  but longer than $2\pi\kappa^{-1}$:
\begin{equation}
    n_0\gamma_m \ll \frac{2\pi}{\tau} \ll \kappa \ll \Omega_m\\
\end{equation}
This allows us to neglect all terms rotating faster than $\Omega_m$ in the temporal enveloppe of the cavity field, to ignore the cavity transient behavior, and to neglect mechanical damping  \cite{Woolley13,Galland14}. Since in the experiment photons are filtered around $\omega_c$ at the resonator output; we focus below on the component $\hat{a}$ of the cavity operator evolving close to the cavity frequency \cite{Woolley13,Galland14}, which is linked to the input and output of the cavity through $\hat{a}_{out}=-\hat{a}_{in}+\sqrt{\kappa}\hat{a}$ . Neglecting the transient dynamics, the following relations are then obeyed:
\begin{align}
\label{equ:Langevin1}
    \hat{a}_{_\pm,out}&=\hat{a}_{in}+i\sqrt{2\tilde{g}_\pm}\hat{b}_{\pm}^{\dag,}\\
\label{equ:Langevin2}
    \frac{d\hat{b}_{\pm}}{dt} &= \pm \tilde{g}_\pm \hat{b}_{\pm} + i\sqrt{2\tilde{g}_\pm}\hat{a}_{in}^{\dag,}
\end{align}
under the condition of weak interaction  $g_0  \ll \kappa, \omega_m$. In the terms on the right, note the implication of $\hat{b}$, $\hat{a}_{in}$ for the drive - and of their adjoints for the drive +. $\tilde{g}_\pm= 2g_0^2 n_{c_\pm} /\kappa$ with $n_{c_\pm} =\frac{\kappa P_\pm}{\hbar\omega_c(\Omega_m^2+\kappa^2/4)}$ the number of intracavity photons at frequency $\omega_\pm$. 



With these approximate relations, we can now model the effect of blue and red-detuned pulses. A blue drive pulse of duration $T_+$ with weak power $P_+$ implements a squeezing operator that creates photon-phonon pairs in equal proportion, while the subsequent red pulse of duration $T_-$ and weak power $P_-$ implements a beamsplitter-like interaction converting phonons into photons at frequency $\omega_c$, allowing the state transfer from mechanics to optics. In order to describe these two sequences, we introduce the following temporal modes already considered in \cite{Hofer11,Vivoli16,Rakhubovsky19}:

\begin{subequations}
\begin{align}
\label{equ:Temporal1}
    \hat{A}_{\pm,in}(t) = \sqrt{\frac{2\tilde{g}_\pm}{e^{\mp\tilde{g}_\pm t}(e^{\tilde{g}_\pm t}-e^{-\tilde{g}_\pm t})}}
    \int_0^t{dt^\prime e^{\mp\tilde{g}_\pm t^\prime}\hat{a}_{in}(t^\prime)}\\
\label{equ:Temporal2} 
    \hat{A}_{\pm,out}(t) = \sqrt{\frac{2\tilde{g}_\pm}{e^{\pm\tilde{g}_\pm t}(e^{\tilde{g}_\pm t}-e^{-\tilde{g}_\pm t})}}
    \int_0^t{dt^\prime e^{\pm\tilde{g}_\pm t^\prime}\hat{a}_{\pm,out}(t^\prime)}
\end{align}
\end{subequations}

and look for evolution operators $\hat{U}_\pm(T_\pm)$ that satisfy $\hat{A}_{\pm,out}(T_\pm) = \hat{U}^\dag_\pm(T_\pm) \hat{A}_{\pm,in}(T_\pm) \hat{U}_\pm(T_\pm)$ and $\hat{b}(T_\pm) = \hat{U}_\pm^\dag(T_\pm)\hat{b}(0)\hat{U}_\pm(T_\pm)$. As detailed in Appendix \ref{app:propag}, we find as a solution for the blue-pulse:
\begin{equation}
\hspace{-1,8cm}
    \hat{U}_+(T_+) = e^{i\tanh q\hat{A}_{+,in}^\dag\hat{b}^\dag(0)}(\cosh q)^{-1-\hat{A}_{+,in}^\dag\hat{A}_{+,in}-\hat{b}^\dag(0)\hat{b}(0)}e^{{i\tanh q\hat{A}_{+,in}\hat{b}(0)}}\\
\end{equation}
with $\cosh q = \exp(\tilde{g}_+T_+)$, an expression that differs from that of \cite{Galland14}. For the red pulse we obtain:
\begin{align}
\hspace{-1cm}
    \hat{U}_-(T_-) = e^{i\tan r\hat{A}_{-,in}^\dag\hat{b}(0)}(\cos r)^{-\hat{A}_{-,in}^\dag\hat{A}_{-,in}+\hat{b}^\dag(0)\hat{b}(0)}e^{{i\tan r\hat{A}_{-,in}\hat{b}^\dag(0)}}
\end{align}
with $\cos r = \exp(-\tilde{g}_-T_-)$, an expression that coincides with that found in \cite{Vivoli16}, up to a different ordering choice (see Appendix \ref{app:propag}).

\subsubsection{Application to our protocol}

We consider now the experimental setup described in Fig.~\ref{fig:scheme}, featuring two identical optomechanical resonators $\mathrm{OM}_i$, $i \in \{1,2\}$ whose mechanical degree of freedom is initially in a thermal state with equal average occupancy $n_{01} = n_{02} = (e^{\hbar \Omega_m /k_B T}-1)^{-1} \ll 1$. Possible differences between the two resonators and their consequences on the performances of the present protocol are discussed in the Appendix~\ref{app:diffreson}. Since we start in the vacuum state for both optical modes, the initial total density matrix is 
\begin{align}
    \rho_{tot,in}& = \ket{00}\bra{00}_{A_1A_2}\otimes \rho^{th}_{b_1}(n_0)\otimes \rho^{th}_{b_2}(n_0)
\end{align}

where we adopt back a Schr\"odinger picture (but rotated by $H_0$) and where $\hat{A}_i=\hat{A}_{i,in}(T)$ and $\hat{b}_i=\hat{b}_i(0)$. The first step of the protocol consists in driving the devices with a blue pulse that generates a phonon delocalized over the two mechanical resonators \cite{Ried18}.
We apply the evolution operator for a blue pulse of duration $T_+$ of strength $\tilde{g}_+$ (tunable \textit{via} laser power) and obtain the density matrix at the end of the interaction: 
\begin{subequations}
\begin{align}
    \rho_{tot,out} = \hat{U}_{+}^{tot}(T_+)\rho_{tot,in}\hat{U}_{+}^{tot,\dag}(T_+)\\
    \hat{U}_{+}^{tot} = \hat{U}_+^{(A_1b_1)}\otimes \hat{U}_+^{(A_2b_2)}
\end{align}
\end{subequations}

Note that postselecting on the presence of a photon at the interferometer output
and tracing on the optical modes would produce a mechanical density matrix close to $\ket{01}\bra{01}_{b_1b_2} + \ket{10}\bra{10}_{b_1b_2}$, such as investigated elsewhere \cite{Ried18}. Here we instead perform a multiple-photon interference with the optical state to be teleported $\ket{\Phi} = \alpha \ket{H} + \beta\ket{V}$ by applying a perfect 50:50 beamsplitter operator $\hat{U}_{BS}$ and postselecting the events where at least one photon goes out of each port. At this stage, the teleportation \textbf{(1)} is performed. We can compute the fidelity $\mathcal{F}_+$ of the mechanical state with respect to $\ket{\Psi^{rot}_{mech}}$ as well as the probability $p_+$ for such a coincidence event to happen.
For the subsequent readout of the teleported state \textbf{(2)}, we apply a red-detuned pulse of duration $T_-$ with interaction strength $\tilde{g}_-$ and trace out the mechanical modes, in cases where one photon a least was detected. Similarly, we compute the fidelity $\mathcal{F}_-$ of the obtained optical density matrix with respect to the ideal rotated state $\ket{\Psi^{rot}_{opt}} = \beta\ket{H}-\alpha\ket{V}=-\sigma_x\sigma_z \ket{\Phi}$, as well as the probability $p_-$ that a photon was indeed detected.

\section{Numerical simulations}

We compute numerically the fidelity and probability of success for the two steps \textbf{(1)} and \textbf{(2)} of the protocol, as a function of the initial occupancy $n_0$ and of the interaction strengths $\tilde{g}_+T_+$ and $\tilde{g}_-T_-$ for the blue and red pulses. Unless explicitely specified, we work with real and equal amplitudes $\alpha = \beta= 1/\sqrt{2}$. 

\subsubsection{Dependence on the initial phonon occupancy}

\begin{figure*}[ht!]
    \includegraphics[width=1\textwidth]{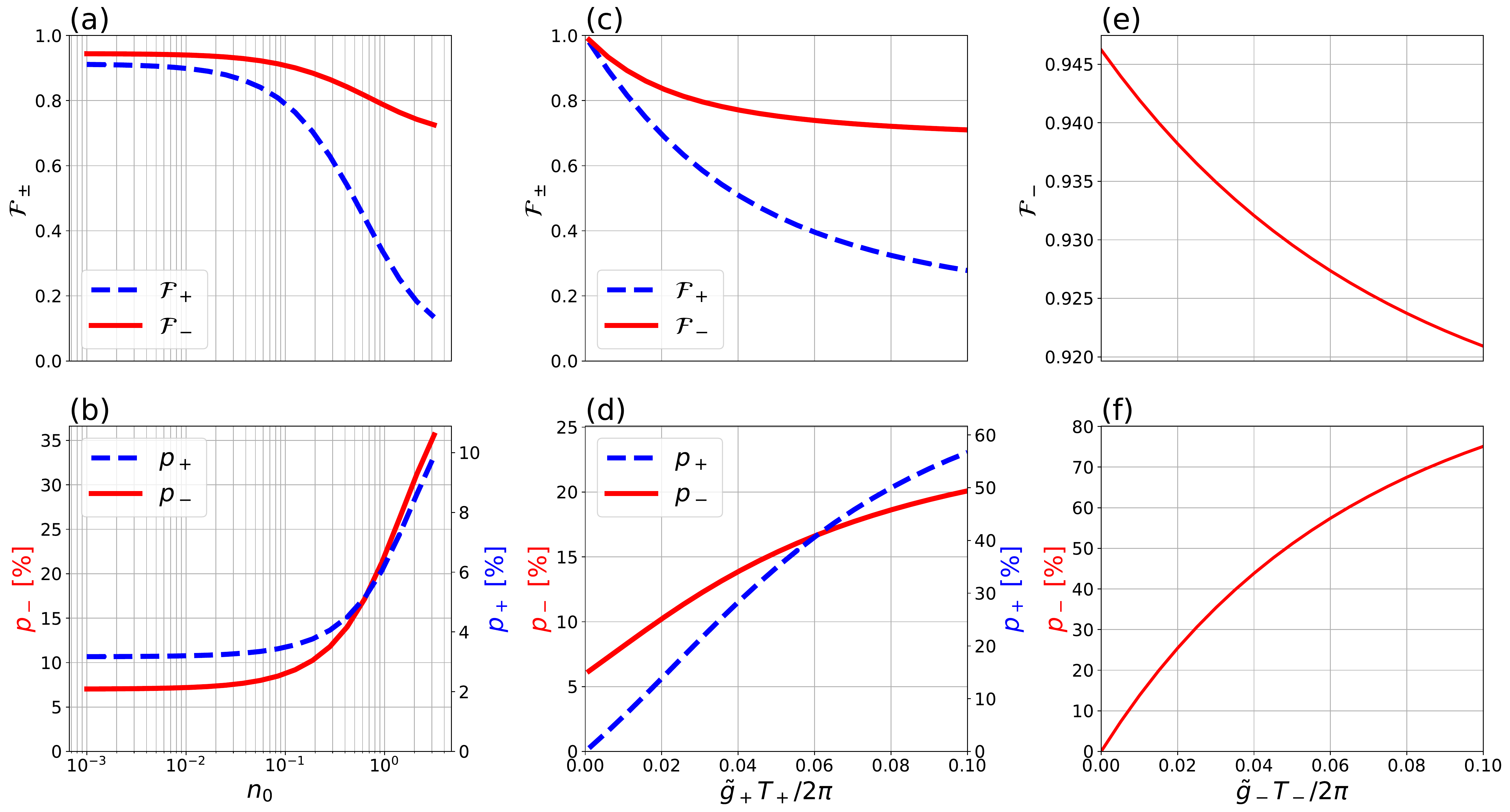}
    \caption{(a) Fidelities $\mathcal{F}_\pm$ and (b) probabilities of success $p_\pm$ after the two steps of the protocol as a function of the initial phonon occupancy, for a fixed $\tilde{g}_\pm T_\pm/2\pi$ of $ 4.9\cdot10^{-3}$. 
    Blue dashed lines: $\mathcal{F}_+$ of the dual-rail-encoded phononic state with respect to the rotated ideal state $\ket{\Psi^{rot}_{mech}}$, and probability $p_+$ of successful coincidence event; red solid lines: $\mathcal{F}_-$ of the photon state obtained after state transfer with respect to $\ket{\Psi^{rot}_{opt}}$, and probablity $p_-$.
    (c) Fidelities $\mathcal{F}_\pm$ and (d) probabilities of success $p_\pm$ as a function of $\tilde{g}_+ T_+/2\pi$ while $\tilde{g}_- T_-/2\pi$ is set to $ 4.9\cdot10^{-3}$ and $n_0$ to $10^{-3}$.
    (e) Fidelity $\mathcal{F}_-$ and (d) probability $p_-$ as a function of $\tilde{g}_- T_-/2\pi$ while $\tilde{g}_+ T_+/2\pi$ is set to $ 4.9\cdot10^{-3}$ and $n_0$ to $10^{-3}$.}
    \label{fig:F_allinone}
\end{figure*}

The results shown in Fig. \ref{fig:F_allinone} (a) and (b) for $\tilde{g}_+T_+/2\pi=\tilde{g}_-T_-/2\pi = 4.9\cdot10^{-3}$ feature a clear deterioration of the fidelity as $n_0$ increases, for both steps \textbf{(1)} and \textbf{(2)}. Indeed, a larger initial thermal population of phonons implies a less pure ground state, an error that propagates to the mechanical state obtained after teleportation. The counterpart of this degraded mechanical state is that the probability to obtain coincidences events after the blue pulse ($p_+$) and to obtain a photon after the red pulse ($p_-$) also increase, equally drastically. In other words the probability of the protocol to be completed is increased, but at the expense of its fidelity. The increase of $p_+$ can be understood by an approximate analytical treatment (see Appendix \ref{app:calcul}), while a larger $n_0$ also implies that a larger number of phonons are available for annihilation by the red pulse, explaining the increase in the probability $p_-$. 

The fidelity $\mathcal{F}_-$ after step \textbf{(2)} is larger than the mechanical fidelity $\mathcal{F}_+$ after step \textbf{(1)}, a result that may appear counter-intuitive at first. Indeed, at the end of step \textbf{(1)}, the obtained mechanical density matrix has multiphonon terms because of non-zero initial phonon occupancy. Its mapping to an optical density matrix in step \textbf{(2)} is not faithful since postselection on single anti-Stokes photons at $\omega_c$ is performed at that stage. This reduction of the relevant Hilbert space is accompanied by the fact that these single photons do not carry the whole information formerly encoded mechanically, letting instead single and multiphonon terms in the mechanical density matrix lead to indistinguishable optical contributions. This purifies the obtained optical state with respect to the mechanical state it was mapped from. The situation is different for the considered mechanical Hilbert space, whose dimensions must be large enough to cover all the significantly populated Fock states, and where the presence of multiphonon terms decrease the fidelity to the target mechanical state.

\subsubsection{Dependence on interaction strengths}

We now choose an initial phonon occupancy of $10^{-3}$ in order to operate in a regime of large fidelity, a reachable regime with very high mechanical frequency resonators and millikelvin temperatures, and set $\tilde{g}_-T_-/2\pi = 4.9\cdot10^{-3}$ while we sweep $\tilde{g}_+T_+$. Conversely, we set $\tilde{g}_+T_+/2\pi = 4.9\cdot10^{-3}$ and sweep $\tilde{g}_-T_-$. The results are shown on Fig.~\ref{fig:F_allinone} (c) to (f).

As expected, a too strong blue pulse triggers multiple phonon generation that deteriorates the fidelity, while increasing the probability of success of the first step.  The results of Fig.\ref{fig:F_allinone} (e) and (f) show that we can increase the red pulse strength in order to increase the probability of success of the second step, while conserving a decent associated fidelity. This is of course in the ideal case in absence of transient heating during the pulse, which would populate higher phonon Fock states \cite{Meenehan15}. In such case, a strong red-detuned pulse could annihilate more than one phonon, inducing random multiphoton states before the optical tomography. Note that the requirement of small $\tilde{g}_\pm T_\pm$, combined with $T_\pm \gg 2\pi\kappa^{-1}$ translates into the experimentally relevant weak coupling regime $g_0\sqrt{n_{c\pm}} \ll \kappa$.

\subsubsection{Experimental feasibility}
\label{section:expfeas}

Let us now discuss the feasibility of this scheme with real experimental parameters. We consider the following device and pulse parameters, compatible with recent experimental realizations \cite{Ried18,Marinkovic18,Wall19}:

\begin{center}
    \begin{tabular}{c|c|c|c|c|c}
        $\Omega_m/2\pi$ [GHz] & $T$ [mK] & $\kappa_e/2\pi$ [MHz] & $g_0/2\pi$ [kHz] & $T_\pm$ [ns] & $n_{c\pm}$\\
        \hline
        5.0 & 20 & 200 & 700 & 10 & 100\\
    \end{tabular}
\end{center}
which lead to the three following dimensionless parameters relevant to the protocol:
\begin{center}
    \begin{tabular}{c|c|c}
        $n_0$ & $\tilde{g}_+T_+/2\pi$ & $\tilde{g}_-T_-/2\pi$\\
        \hline
        $6\cdot10^{-6}$ & $4.9\cdot10^{-3}$ & $4.9\cdot10^{-3}$\\
    \end{tabular}
\end{center}
and to the following expectations for the fidelities and probabilities of success of the two steps:

\begin{center}
    \begin{tabular}{c|c|c|c|c}
        
        $\mathcal{F}_+$ & $\mathcal{F}_-$ & $p_+$ [\%] & $p_-$ [\%]& $p_+p_-$[\%]\\
        \hline
        0.91 & 0.94 & 3.2 & 7.0 & 0.22\\
    \end{tabular}    
    
\end{center}
The condition that the pulse duration is longer than $2\pi\kappa^{-1}$ is respected by a factor of only five in this example, underlining the need for good optical cavities. At equal interaction strength $\tilde{g}T$, the pulse can be made longer at the expense of a lower intracavity photon number, which will degrade the signal to noise ratio (See Appendix.\ref{app:nNEP}). Still, these results show that our protocol can lead to relatively high fidelities already with realistic parameters.

\section{Conclusion}
We have described an optomechanical discrete variable quantum teleportation scheme that is readily implementable with state-of-the-art optomechanical devices. Based on a heralded probabilistic principle, this protocol enables teleportation, mechanical storage and further readout of an arbitrary qubit state originally encoded on a single photon. The protocol is expected to allow high fidelity, not only for the storage of the qubit in the mechanical memory ($\mathcal{F}_+$), but also for the retrieval of the single-photon flying qubit at the output, with an expected fidelity $\mathcal{F}_-$ reaching 0.94. These features pave the way for future experiments in the realm of quantum communications and memories with optomechanics, a technology that is yet to be explored in this context.

\subsubsection{Acknowledgments}
The authors thank Cristiano Ciuti for useful discussions and acknowledge support by the European Research Council through the ERC NOMLI project, and of the R\'egion Ile de France through the SIRTEQ OMQuTe project.

\bibliography{apssamp}

\nocite{*}

\appendix

\begin{widetext}

\section{Evolution operators of our model}
\label{app:propag}

Starting from the Langevin equations (Eqs. \ref{equ:Langevin1} and \ref{equ:Langevin2}) and from the definition of the temporal mode (Eqs. \ref{equ:Temporal1} and \ref{equ:Temporal2}), whose exponential shape is an approximation valid in the adiabatic regime $g_0 \ll \kappa$, we obtain below the evolution operators $ \hat{U}$. By integrating Eqs. \ref{equ:Langevin1} and \ref{equ:Langevin2}, we re-express $\hat{A}_{\pm,out}$ and $\hat{b}_\pm$: 
\begin{align}
\label{Bogo}
    \hat{A}_{\pm,out}(t) &= e^{\pm\tilde{g}_\pm t} \hat{A}_{\pm,in}(t)+ i\sqrt{e^{\pm\tilde{g}_\pm t}(e^{\tilde{g}_\pm t}-e^{-\tilde{g}_\pm t})}\hat{b}^{\dag,}(0)\\
\label{Bogo2}
    \hat{b}_\pm(t) &= e^{\pm \tilde{g}_\pm t}\hat{b}(0) + i\sqrt{e^{\pm\tilde{g}_\pm t}(e^{\tilde{g}_\pm t}-e^{-\tilde{g}_\pm t})}\hat{A}_{\pm,in}^{\dag,}(t)
\end{align}

The procedure to express the evolution operator for the blue pulse \textbf{(1)} obeying $\hat{A}_{+,out}(t) = \hat{U}^\dag_+(t) \hat{A}_{+,in}(t) \hat{U}_+(t)$ and $\hat{b}_+(t) = \hat{U}_+^\dag(t)\hat{b}(0)\hat{U}_+(t)$ follows closely a work from 1988 by Truax \cite{Truax88}. By defining $\cosh(q)=e^{\tilde{g}_+t}$ and $\sinh(q) = \sqrt{e^{2\tilde{g}_+t}-1}$, we show that the unitary operator $\hat{U}_+(t)=\exp[iq(\hat{A}_{+,in}(t)\hat{b}(0)+h.c)]$ realizes the desired rotation.
By writing $\hat{U}^\dag_+(t) = \exp\hat{B}$, we have indeed:
\begin{align}
    \hat{A}_{+,out}(t) &= e^{\hat{B}}\hat{A}_{+,in}(t) e^{-\hat{B}} = \sum_{k\in \mathbb{N}}{\frac{1}{k!}\underbrace{[\hat{B},[\hat{B},...,[\hat{B},\hat{A}_{+,in}(t)]]]}_{k~\text{times}}}\\
    &= \cosh(q)\hat{A}_{+,in}(t) + i\sinh(q) \hat{b}^\dag(0)
\end{align}
where we used that the nested commutators equal $q^k \hat{A}_{+,in}(t)$ for $k$ even and $iq^k\hat{b}^\dag(0)$ for $k$ odd. Similarly:
\begin{equation}
    \hat{b}_+(t) = \cosh(q)\hat{b}(0) + i \sinh(q)\hat{A}_{+,in}^\dag(t)
\end{equation}
retrieving indeed Eqs. \ref{Bogo} and \ref{Bogo2}.

We express now $\hat{U}_+(t)$ in a more convenient manner. We note that $\hat{U}_+(t) = \exp[iq(\hat{L}_{+}+\hat{L}_{-})]$ with $\hat{L}_{+} = \hat{L}^\dag_{-} = \hat{A}^\dag_{+,in}(t)\hat{b}^\dag(0)$, such that by defining $\hat{L}_0=\hat{L}^\dag_0= \frac{1}{2}(1+\hat{A}_{+,in}^\dag(t)\hat{A}_{+,in}(t)+\hat{b}_+^\dag(0)\hat{b}_+(0))$, we obtain three generators of the $\mathfrak{su}(1,1)$ Lie algebra ($[\hat{L}_{-},\hat{L}_{+}]=2{L}_0,[\hat{L}_0,\hat{L}_\pm]=\pm{L}_\pm$). From Eq. (23b) of Ref.\cite{Truax88}, one obtains the evolution operator for the blue pulse:
\begin{equation}
    \hat{U}_+(t) = e^{i\tanh q\hat{A}_{+,in}^\dag\hat{b}^\dag(0)}(\cosh q)^{-1-\hat{A}_{+,in}^\dag\hat{A}_{+,in}-\hat{b}^\dag(0)\hat{b}(0)}e^{{i\tanh q\hat{A}_{+,in}\hat{b}(0)}}\\
\end{equation}

The procedure to obtain the red pulse \textbf{(2)} evolution operator is similar and starts by defining $\cos(r) = e^{-\tilde{g}_-t}$ and $\sin(r) = \sqrt{1-e^{-2\tilde{g}_- t}}$. Let us show that $\hat{U}_-(t) = \exp[ir(\hat{A}^\dag_{-,in}(t)\hat{b}(0)+h.c)]$ realizes the desired rotation $\hat{A}_{-,out}(t) = \hat{U}^\dag_-(t) \hat{A}_{-,in}(t) \hat{U}_-(t)$ and $\hat{b}_-(t) = \hat{U}_-^\dag(t)\hat{b}(0)\hat{U}_-(t)$. Writing $\hat{U}^\dag_-(t)=\exp\hat{B}$, we have:
\begin{align}
    \hat{A}_{-,out}(t) &= e^{\hat{B}}\hat{A}_{-,in}(t) e^{-\hat{B}} = \sum_{k\in \mathbb{N}}{\frac{1}{k!}\underbrace{[\hat{B},[\hat{B},...,[\hat{B},\hat{A}_{-,in}(t)]]]}_{k~\text{times}}}\\
    &=\sum_{k\in \mathbb{N}}{\left[\frac{(-1)^k}{(2k)!}r^{2k}\hat{A}_{-,in}(t) +i \frac{(-1)^k}{(2k+1)!}r^{2k+1}\hat{b}(0) \right]}\\
    &= \cos(r)\hat{A}_{-,in}(t) + i\sin(r) \hat{b}(0)\\
    \hat{b}_-(t) &= \cos(r)\hat{b}(0) + i \sin(r)\hat{A}_{-,in}(t)
\end{align}

Again, $\hat{U}_-(t) =\exp[ir(\hat{L}_{+}+\hat{L}_{-})]$ but this time the operators $\hat{L}_{+} = \hat{L}^\dag_{-} = \hat{A}_{-,in}^\dag(t)\hat{b}_-(0)$, $\hat{L}_0=\hat{L}^\dag_0= \frac{1}{2}(\hat{A}_{-,in}^\dag(t)\hat{A}_{-,in}(t)-\hat{b}(0)\hat{b}^\dag(0))$ are three generators of the $\mathfrak{su}(2)$ Lie algebra ($[\hat{L}_{-},\hat{L}_{+}] = -2\hat{L}_0, [\hat{L}_0,\hat{L}_{\pm}]= \pm \hat{L}_{\pm}$). Then from Eq. (23a) of Ref.~\cite{Truax88}, one has:
\begin{equation}
    \hat{U}_-(t) = e^{i\tan r\hat{A}_{-,in}^\dag\hat{b}(0)}(\cos r)^{-\hat{A}_{-,in}^\dag\hat{A}_{-,in}+\hat{b}^\dag(0)\hat{b}(0)}e^{{i\tan r\hat{A}_{-,in}\hat{b}^\dag(0)}}
\end{equation}

\section{Approached analytical calculation}
\label{app:calcul}

We develop an approached analytical description of the proposed protocol, providing an explicit expression of the mechanical density matrix after the first pulse in the simplified case where the detectors are photon-number resolving. We compute the probability for a coincidence event to happen, adding explanations to the results of Fig. \ref{fig:F_allinone}. We then treat the red pulse and describe the contribution of dominant terms.

\subsubsection{Blue pulse}
We drop the $+$ in this blue-detuned section. We start from the initial total optomechanical density matrix describing the two mechanical modes $\hat{b}_i$ as well as the two optical modes $\hat{A}_i$:

\begin{align}
    \rho_{tot,in}&= \ket{0}\bra{0}_{\hat{A}_1}\otimes\ket{0}\bra{0}_{\hat{A}_2}\otimes \rho_{b,1}^{th}(n_0) \otimes \rho_{b,2}^{th}(n_0)\\
    \rho_{b,i}^{th}(n_0) &= (1-p_i)\sum_{n_i}p_i^{n_i}\ket{n_i}\bra{n_i} , p_i = n_{0i}/(1+n_{0i})
\end{align}

and apply to each OM resonator the blue pulse evolution operator of duration $T$, allowing for various $q_i$ and $n_{0i}$.
\begin{equation}
    \rho_{tot,out}=\frac{(1-p_1)(1-p_2)}{(\cosh q_1 \cosh q_2)^2}\sum_{n_1,n_2,l_1,l_2,l_1',l_2'} \bar{p_1}^{n_1}\bar{p_2}^{n_2} \Lambda(n_1,n_2,l_1,l_2,l_1',l_2') \ket{l_1,l_2,n_1+l_1,n_2+l_2}\bra{l_1',l_2',n_1+l_1',n_2+l_2'}
\end{equation}

~~\\with $\Lambda(n_1,n_2,l_1,l_2,l_1',l_2') = (i\tanh q_1)^{l_1}(i\tanh q_2)^{l_2}(-i\tanh q_1)^{l_1'}(-i\tanh q_2)^{l_2'}
\sqrt{\frac{(n_1+l_1)!}{l_1!n_1!}}\sqrt{\frac{(n_2+l_2)!}{l_2!n_2!}}\sqrt{\frac{(n_1+l_1')!}{l_1'!n_1!}}\sqrt{\frac{(n_2+l_2')!}{l_2'!n_2!}}$ and $\bar{p_i} = p_i\cosh q_i^{-2}$.$\ket{l_1,l_2,n_1,n_2}$ is the state with $l_i$ photons at $\omega_c$ and $n_i$ phonons at $\Omega_m$ in the resonator $\text{OM}_i$.\\

After conditioning on the detection of one (and only one) photon with the projector $C = (\ket{01}\bra{01}_{A_1A_2}+\ket{10}\bra{10}_{A_1A_2})\otimes \mathbb{I}_{b_1b_2}$ and tracing on the optical modes, we obtain the normalized bipartite mechanical state:


\begin{align}
\rho_{b_1b_2}^{cond} = (1-\bar{p_1})(1-\bar{p_2})&\sum \bar{p_1}^{n_1}\bar{p_2}^{n_2}\left(\frac{\tanh^2 q_1}{(1-\bar{p_1})}+\frac{\tanh^2q_2}{(1-\bar{p_2})}\right)^{-1} \nonumber\\
&[\tanh^2q_1(n_1+1)\ket{n_1+1,n_2}\bra{n_1+1,n_2}+\tanh^2q_2(n_2+1)\ket{n_1,n_2+1}\bra{n_1,n_2+1}]
\end{align}
This last state corresponds to the situation investigated in Ref.\cite{Ried18}. We now treat the Bell state measurement. We restart from the unconditioned total density matrix after blue pulse and tensor product it with the to-be-teleported density matrix $\rho_c = \begin{pmatrix}
|\alpha|^2 & \alpha\beta^* \\
\alpha^*\beta & |\beta|^2
\end{pmatrix}$
describing a photonic qubit in the $H/V$ basis.


In order to condition on coincidences, \textit{i.e.} on the the Bell state $\ket{\Psi^-}=\frac{1}{\sqrt{2}}(\ket{HV}-\ket{VH}) \equiv \frac{1}{\sqrt{2}}(\ket{H01}-\ket{V10})$ we employ the following conditioning operator, which acts before the Bell measurement beamsplitter: 
\begin{equation}
    \hat{K} = \frac{1}{2}\left[\ket{H01}\bra{H01}+\ket{V10}\bra{V10}-\ket{H01}\bra{V10}-\ket{V10}\bra{H01}\right]\otimes \mathbb{I}_{b_1b_2}
\end{equation}

After tracing on the optical modes, we obtain the unnormalized mechanical density matrix conditioned to a successful coincidence event: 
\begin{equation}
    \rho^{cond,coin}=\frac{(1-p_1)(1-p_2)}{2(\cosh q_1 \cosh q_2)^2}\sum_{n_1,n_2}{\bar{p_1}^{n_1} \bar{p_2}^{n_2} \ket{\Psi(n_1,n_2)}\bra{\Psi(n_1,n_2)}}
\end{equation}
with $\ket{\Psi(n_1,n_2)} = \beta \tanh q_1 \sqrt{n_1+1}\ket{n_1+1,n_2} - \alpha \tanh q_2 \sqrt{n_2+1} \ket{n_1,n_2+1}$.
The matrix must then be normalized: 
\begin{align}
Tr[\rho^{cond,coin}] &= \Xi \Gamma~~~~
\mathrm{ with }~~~~ \Xi=\frac{(1-p_1)(1-p_2)}{2(\cosh q_1 \cosh q_2)^2}  ~~~~\mathrm{ and }~~~~\Gamma = \frac{|\beta|^2\tanh^2q_1}{(1-\bar{p_1})^2(1-\bar{p_2})} + \frac{|\alpha|^2\tanh^2q_2}{(1-\bar{p_2})^2(1-\bar{p_1})}
\end{align}
Setting $q_1=q_2$ and $p_1=p_2$, we observe that the product $\Xi\Gamma$ is an increasing function of $p$, hence of $n_0$, which explains the results of Fig. \ref{fig:F_allinone}. 
Finally, the normalized mechanical density matrix is:
\begin{equation}
\label{eq:rho_mech_cond}
\rho^{cond,coin,*}= \Gamma^{-1}\sum_{n_1,n_2}{\bar{p_1}^{n_1} \bar{p_2}^{n_2} \ket{\Psi(n_1,n_2)}\bra{\Psi(n_1,n_2)}}
\end{equation}

\subsubsection{Red pulse}

For the red pulse, we compute the effect of the evolution operator on a generic mechanical state with the optical modes in the vacuum state:
\begin{align}
    \hat{U}_{-}^{tot}\ket{0,0,n_1,n_2} &= \sum\limits_{l_1\leq n_1, l_2\leq n_2} F(l_1,l_2,n_1,n_2) \ket{l_1, l_2, n_1-l_1,n_2-l_2}\\
    F(l_1,l_2,n_1,n_2) &= \cos r_1 ^{n_1} \cos r_2 ^{n_2} (i \tan r_1)^{l_1} (i \tan r_2)^{l_2}\sqrt{\frac{n_1!}{(n_1-l_1)!l_1!}}\sqrt{\frac{n_2!}{(n_2-l_2)!l_2!}}
\end{align}
Such that the generic term in a density matrix evolves as:
\begin{align}
&\hat{U}_-^{tot}\ket{0,0,n_1,n_2}\bra{0,0,n_1',n_2'}\hat{U}_-^{tot,\dag} = \\
&\sum\limits_{l_1\leq n_1, l_2\leq n_2} 
\sum\limits_{l_1'\leq n_1', l_2'\leq n_2'} F(l_1,l_2,n_1,n_2)F(l_1',l_2',n_1',n_2')^*
\ket{l_1, l_2, n_1-l_1,n_2-l_2}\bra{l_1', l_2', n_1'-l_1',n_2'-l_2'}
\end{align}
\\
\\
We apply this formula to the four terms of $\ket{0,0}\otimes\ket{\Psi(n_1,n_2)}\bra{\Psi(n_1,n_2)}\otimes\bra{0,0}$ for each ($n_1,n_2$) in Eq. \ref{eq:rho_mech_cond} before conditioning on one output photon with the projector $C = (\ket{01}\bra{01}_{A_1A_2}+\ket{10}\bra{10}_{A_1A_2})\otimes \mathbb{I}_{b_1b_2}$. Tracing on the mechanical modes leads the final optical density matrix, which is analyzed by tomography at the end of the protocol. As an example, the action of $\hat{U}_{-}^{tot}$ on the dominant term of Eq. \ref{eq:rho_mech_cond} ($n_1=n_2=0$) finally results in the following optical density matrix:

\begin{align}
Tr_{mech}\left[\hat{U}_{-}^{tot}\ket{0,0}\otimes\ket{\Psi(0,0)}\bra{\Psi(0,0)}\otimes\bra{0,0}\hat{U}_{-}^{tot,\dag}\right] = \nonumber\\
\left\{|F(0,0,1,0)|^2 |\beta|^2 \tanh^2 q_1  + |F(0,0,0,1)|^2 |\alpha|^2 \tanh^2 q_2 \right\}&\ket{00}\bra{00} \nonumber\\
+|F(1,0,1,0)|^2 |\beta|^2 \tanh^2 q_1 &\ket{10}\bra{10} \\
+|F(0,1,0,1)|^2 |\alpha|^2 \tanh^2 q_2 &\ket{01}\bra{01} \nonumber\\
-F(1,0,1,0)F^*(0,1,0,1) \beta\alpha^*\tanh q_1 \tanh q_2&\ket{10}\bra{01} \nonumber\\
-F^*(1,0,1,0)F(0,1,0,1) \beta^*\alpha\tanh q_1 \tanh q_2&\ket{01}\bra{10} \nonumber
\end{align}

We recall that in this notation $\ket{10} \equiv \ket{H}$ and $\ket{01} \equiv \ket{V}$.
Projecting on the non-vacuum state removes the first line and assuming ($r_1=r_2$, $q_1 = q_2$) leaves us with a density matrix proportional to: 
$\begin{pmatrix}
|\beta^2| & -\beta\alpha^* \\
 -\beta^*\alpha & |\alpha|^2
\end{pmatrix}=\sigma_z\sigma_x\rho_c\sigma_x\sigma_z$
in the $H/V$ basis. The calculation for the next term ($n_1=1,n_2=0$) leads to the following result:

\begin{align}
Tr_{mech}\left[\hat{U}_{-}^{tot}\ket{0,0,\Psi(1,0)}\bra{0,0,\Psi(1,0)}\hat{U}_{-}^{tot,\dag}\right] = &\nonumber\\
\left\{2|F(0,0,2,0)|^2|\beta|^2\tanh^2q_1 + |F(0,0,1,1)|^2|\alpha|^2\tanh^2q_2\right\}\ket{00}\bra{00}&\nonumber\\
~\nonumber\\
+\left(2|F(1,0,2,0)|^2|\beta|^2\tanh^2q_1 + \overbrace{|F(1,0,1,1)|^2|\alpha|^2\tanh^2q_2}^{\text{parasitic term}}\right)\ket{10}\bra{10}&\nonumber\\
+ |F(0,1,1,1)|^2|\alpha|^2\tanh^2q_2 \ket{01}\bra{01} \\
-\sqrt{2}\beta\alpha^* \tanh q_1\tanh q_2 F(1,0,2,0)F^*(0,1,1,1)\ket{10}\bra{01}&\nonumber\\
-\sqrt{2}\beta^*\alpha \tanh q_1\tanh q_2 F(0,1,1,1)F^*(1,0,2,0)\ket{01}\bra{10}\nonumber&\\
~\nonumber\\
+ 2|F(2,0,2,0)|^2|\beta|^2\tanh^2q_1 \ket{20}\bra{20} \nonumber\\
+ |F(1,1,1,1)|^2|\alpha|^2\tanh^2q_2 \ket{11}\bra{11} \nonumber\\
- \sqrt{2}\beta\alpha^* \tanh q_1\tanh q_2  F(2,0,2,0)F^*(1,1,1,1)\ket{20}\bra{11} \nonumber\\
- \sqrt{2}\beta^*\alpha \tanh q_1\tanh q_2  F(1,1,1,1)F^*(2,0,2,0)\ket{11}\bra{20} \nonumber
\end{align}

where multiphoton states are present that degrade the computed fidelity. A parasitic term proportional to $|\alpha|^2$ induces a finite matrix element $\ket{10}\bra{10}$ even if $\beta=0$. This makes the fidelity state-dependent as discussed in App. \ref{app:diffreson}.\\

\subsubsection{Realistic projectors}
The above calculation is simplified since we considered ideal photon-number resolving detectors, allowing us to express $\hat{K}$ with only two-photon states. The numerical simulations presented in the main text apply instead a 50:50 beamsplitter interaction before conditioning on coincidence events in the four output modes of the Bell state measurement beamsplitter (2 polarization modes $H/V \times$ 2 spatial modes $L/R$) with the following projector, which acts after the beamsplitter:


\begin{align}
    \hat{K}' = \left( \mathbb{I}_{L_HL_VR_HR_V}- (1-\eta_{detect})^2\ket{0000}\bra{0000} - (1-\eta_{detect})\sum_{(i,j)\neq (0,0)}{\ket{00ij}\bra{00ij}+\ket{ij00}\bra{ij00}}\right)
    \otimes \mathbb{I}_{b_1b_2}
\end{align}
\end{widetext}

where we took into account the detector's dark count rate $\Gamma_{dark}$. Let $\Gamma_{rep}$ be the repetition rate of the protocol, a click on a detector has a probability $\eta_{detect} = \Gamma_{dark}/(\Gamma_{dark}+ \Gamma_{rep})$ to be a false positive. Similarly the right projector to use during the red pulse is $\hat{C}'= [\mathbb{I}_{A_1A_2} - (1-\eta_{detect})\ket{00}\bra{00}]\otimes \mathbb{I}_{b_1b_2}$ in order to allow for multiple-photon states to be treated by the tomography unit with noisy detectors.


\section{Technical challenges}

\subsubsection{Noise equivalent phonon number}
\label{app:nNEP}
\begin{figure}[ht!]
    \centering
    \includegraphics[width=0.5\textwidth]{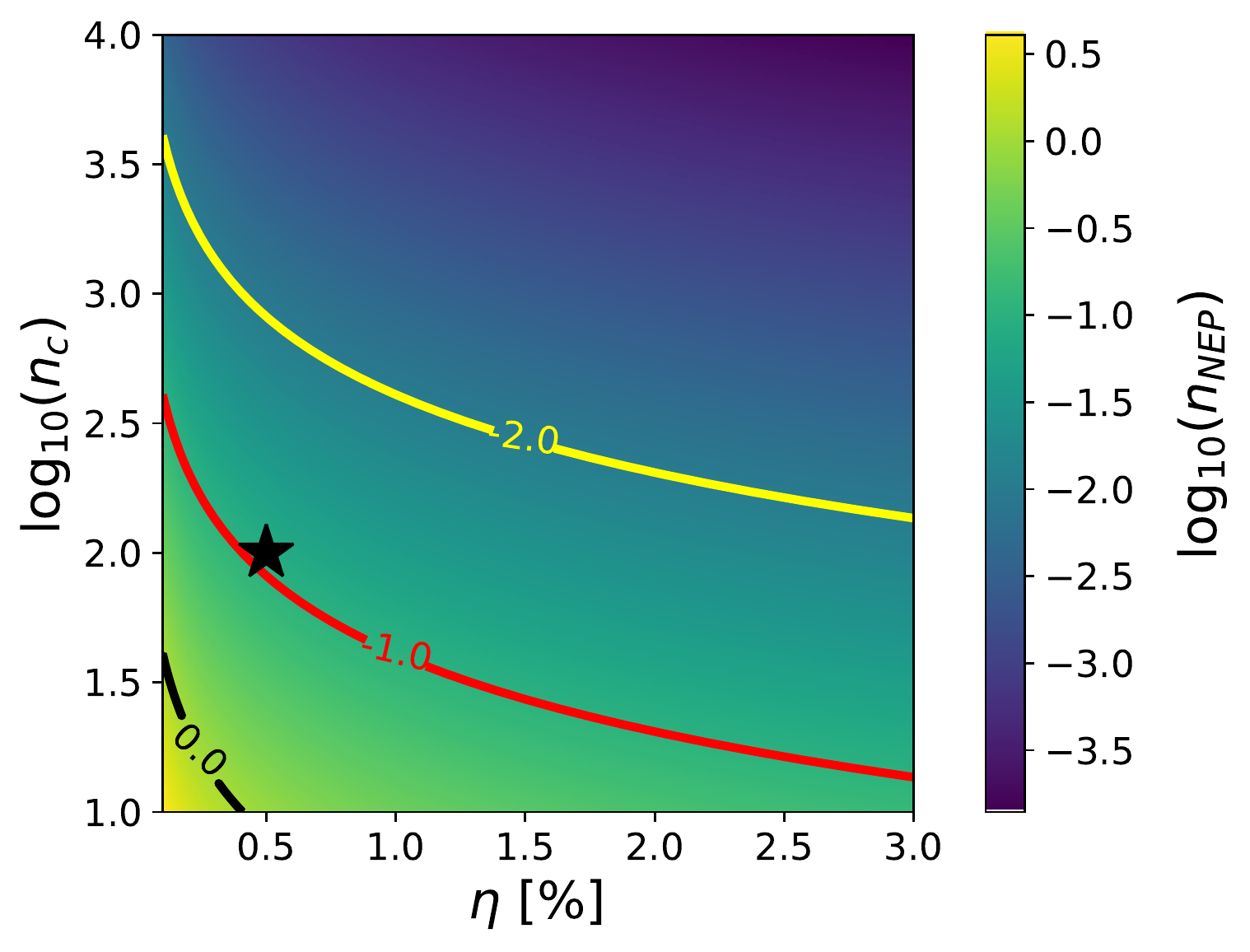}
    \caption{Noise-equivalent phonon number as a function of intracavity photon number $n_c$ and total optical detection efficiency $\eta$. The black star corresponds to the parameters considered in the text and ensures that $n_{NEP} \lesssim 0.1$.}
    \label{fig:nNEP}
\end{figure}

As shown in Fig. \ref{fig:F_allinone} high fidelities are attained for low interaction strengths $\tilde{g}_\pm T_\pm$. For fixed device parameters $g_0$ and $\kappa$, there is hence an upper bound to the product $n_c T_\pm$. In addition a minimal $n_c$ is required, which prevents from arbitrarily increasing $T_\pm$. This is shown by an analysis of the noise-equivalent phonon number $n_{NEP}$, a figure already introduced in past work \cite{Meenehan15}:
\begin{equation}
    n_{NEP} = \frac{\Gamma_{dark}+ \Gamma_{pump}}{\Gamma_{SB}} = \frac{\kappa^2\Gamma_{dark}}{4 \kappa_e \eta g_0^2 n_c} + A\left(\frac{\kappa\Omega_m}{2\kappa_e g_0}\right)^2
\end{equation}
with $\Gamma_{dark}$ the dark-count rate of the photodetectors, $\Gamma_{pump}$ the arrival rate of pump photons that are not properly filtered and leak to the detection, $\Gamma_{SB}$ the sideband photon rate, 
$A$ the attenuation factor from the filter and $\eta$ the total measurement efficiency, including optical losses from the resonator to the detectors and detector efficiency. In Fig.~\ref{fig:nNEP}, we plot $n_{NEP}$ as a function of $n_c$ and $\eta$ for the parameters of section \ref{section:expfeas}, but for the optical coupling, which we assume to be critical $\kappa=2\kappa_e$. We consider $A=-100$$dB$ and $\Gamma_{dark}=100$$Hz$.

With such strong filtering but limited detection efficiency, a value of $n_c\geq 100$ is required to maintain $n_{NEP}$ below $0.1$. Combining this constraint with the bound on $n_cT$ explains why $T_\pm$ cannot be increased arbitrarily, hence our choice of $T_\pm=10~\text{ns}$. 


\subsubsection{Waiting time and probability of success in experiments}
We wish to estimate the minimal waiting time between two successful completions of the protocol, \textit{i.e.} a coincidence event during step \textbf{(1)} followed by a detection event on one of the detectors during step \textbf{(2)}. We impose a maximal duration $T_{off}\leq \gamma_m^{-1}$ between the blue and the red pulse, in order to protect the coherence of the generated mechanical state. But in order to ensure that the devices are at equilibrium with the thermal bath, we also wait at least $T_{relax}$ (typically more than a few $\gamma_m^{-1}$) between two realizations of the protocol.
The time taken to realize the protocol and reinitialize the set-up is then $T_{per} = T_++ T_{off} + T_- + T_{relax}$. The probability of success of the whole protocol (teleportation and read-out) is $p_{success} = (\eta_{sps}\cdot\eta_+p_+ \cdot \eta_-p_-)$, with $\eta_\pm$ the linear efficiencies of the two subsequent optical measurements associated to steps \textbf{(1)} and \textbf{(2)} and $\eta_{sps}$ the efficiency of the single-photon source whose photon-states are to be teleported . 
Since $T_{per}$ is dominated by $T_{relax}$, one can think of reducing the mechanical quality of the devices in order to decrease $T_{per}/p_{success}$, but this will affect their memory capabilities. Since small values of $p_\pm$ are required to ensure large fidelities, it is crucial to develop efficient filtering and optical detection paths. 

\subsubsection{Non-identical resonators}
\label{app:diffreson}
In practice, the two nanofabricated optomechanical devices will present differences in their exact dimensions, typically at the part per thousand level or less \cite{Santos17}, producing a detuning in their optical and mechanical frequencies. Differences in their optical and/or mechanical dissipation can also be sizable. Part of this problem can be compensated by driving in an asymmetric way \cite{Ried18}. One can shift the drive frequency to match the mechanical sideband and balance the drive power in the two arms in order to compensate for different $g_0^2/\kappa$ ratio. However a difference in $\Omega_m$ implies that the two mechanical degrees of freedom precess at different speeds during the off-time between pulses, producing an evolutive dephasing. Moreover the need for indistinguishable photons after scattering in both devices puts a central constraint on their cavity frequency: $|\omega_{c1}-\omega_{c2}|$ should be as low as possible. These aspects can be treated by the use of post-process techniques to reduce the resonator-to-resonators disorder, such as the photo-electrochemical tuning technique demonstrated in nano-optomechanical devices \cite{Santos17}. 

\subsubsection{Dependence on $\alpha$ and $\beta$ }
\begin{figure}[ht!]
    \centering
    \includegraphics[width=0.45\textwidth]{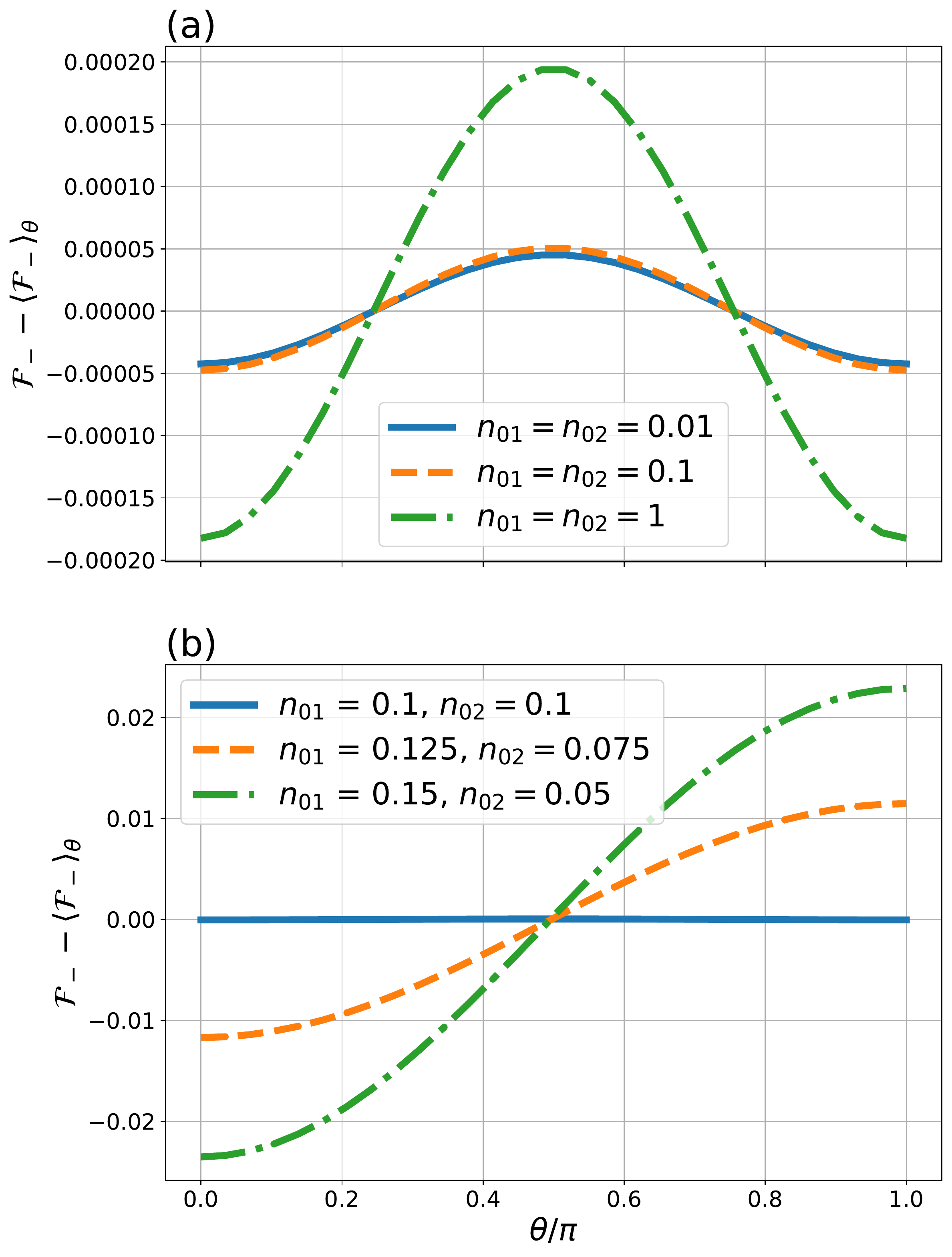}\\
    \caption{Fidelity after the red pulse as a function of the initial state to teleport $\ket{\Phi(\theta, \varphi)} = \cos(\theta/2)\ket{H}+ e^{i\varphi}\sin(\theta/2)\ket{V}$, $\forall \varphi$. For all plots, $\tilde{g}_\pm T_\pm/2\pi = 4.9\cdot 10^{-3}$.}
    \label{fig:fidelVSstate}
\end{figure}

An interesting feature is that all states on the Bloch sphere are not equally handled by the protocol. Indeed, even for equal initial phonon occupancy of the two disks, equatorial states are retrieved with a better fidelity after the red pulse (see Fig.~\ref{fig:fidelVSstate}a). If the state to teleport sits on the pole ($\alpha=1, \beta=0$ for instance), the two resonators must be placed in the phononic state $\ket{01}$. There should be zero horizontal component in the polarization of the output photon after the red pulse, which is not possible since the true phononic occupancy of $\text{OM}_1$ is always larger than 0. In contrast, teleporting an equatorial state $|\alpha| = |\beta|$ is easier because the photon we expect has equal horizontal and vertical polarization components, which is naturally achieved in our balanced Mach-Zehnder configuration. This subtle effect is larger as the mean initial phonon occupancy of the disks grows, and the states that are easier to read-out shift away from the equator as we introduce an imbalance in the initial occupancy of the two disks (see Fig.~\ref{fig:fidelVSstate}b).

\subsubsection{Importance for $\ket{\Phi}$ to be in a single-photon state}
\begin{figure}[ht!]
    \centering
    \includegraphics[width=0.45\textwidth]{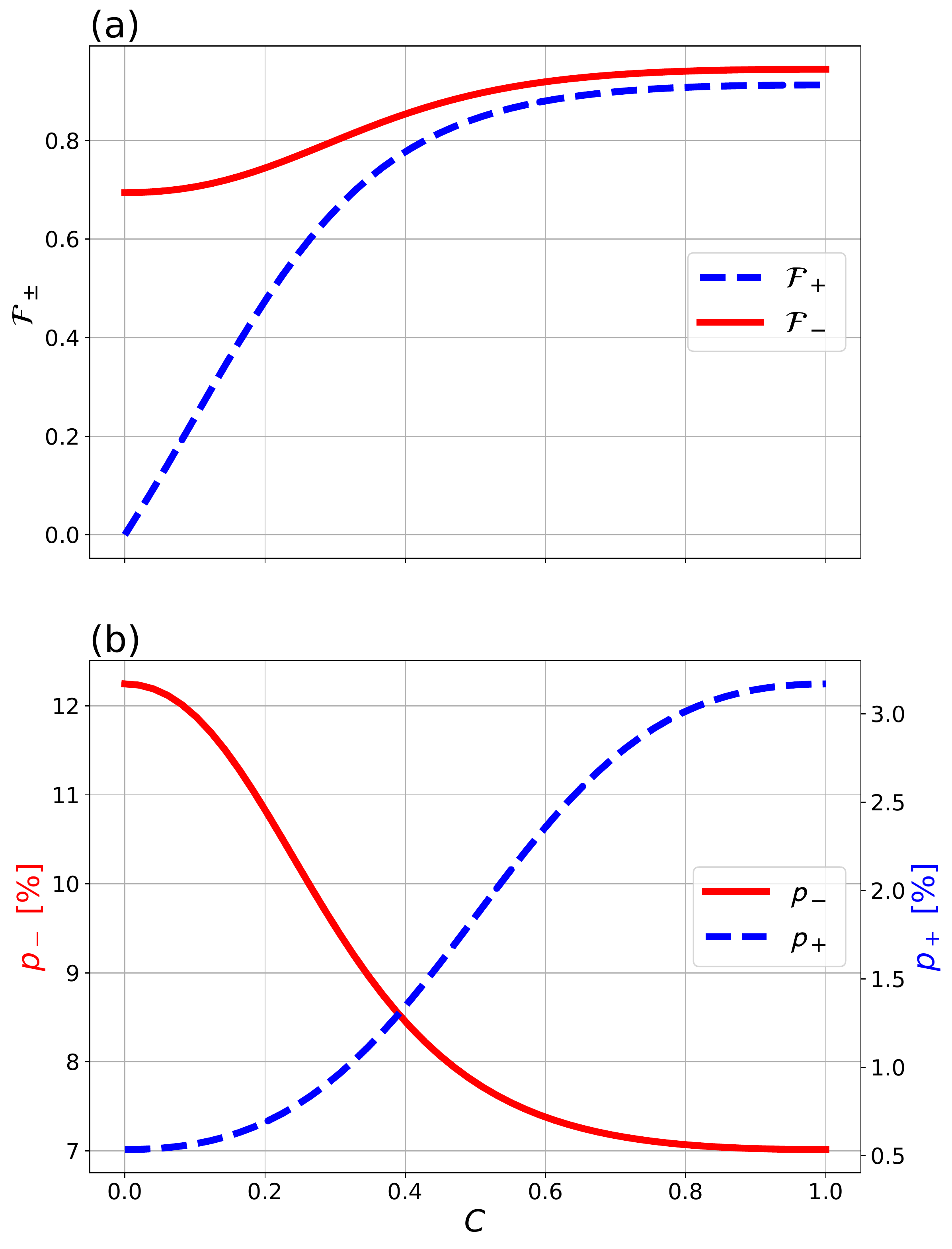}
    \caption{(a) Fidelities and (b) probabilities of success of the two steps of the protocol with respect to $C$. The parameters are those considered in the text.}
    \label{fig:fidelVSC}
\end{figure}

One could imagine using a strongly attenuated coherent source instead of a single-photon source for the Bell measurement, but this would strongly impact the implementation of the protocol. In fact, such attenuated coherent state (with mean photon occupancy $\ll1$) overlaps mostly with the vacuum state, hence produces most of the time no useful coincidence. Additionally a coincidence event in our measurement does not mean necessarily that one photon from the source and one photon from the Mach-Zehnder interferometer were successfully projected in the antisymmetric Bell state $\ket{\Psi^-}$ before the beamsplitter. Coincidence events also take place when a wavepacket containing several photons exits the interferometer and impinges on the Bell measurement beamsplitter, post-selecting undesirable multi-excitation events. This can be appreciated writing $\ket{\Phi(C)} = \sqrt{1-C}\ket{0} + \sqrt{C}(\alpha\ket{H}+ \beta\ket{V})$ and looking at the fidelities and probabilities of success with respect to $C$ in Fig.~\ref{fig:fidelVSC}. As expected, $\mathcal{F}_\pm$ is degraded at low $C$, but $p_-$ increases because events where more than one phonon was created were selected during \textbf{(1)}. This stresses the advantage of using an ideal single-photon source as input of the protocol.

\end{document}